\documentclass[aps,prd,twocolumn,groupedaddress]{revtex4-1}
\usepackage{graphicx}
\include{amssym}

\begin{document}
\title{Decay rates $f_{1}(1285)\to \rho^{0}\pi^{+} \pi^{-}$ and $a_{1}(1260)\to\omega\pi^{+}\pi^{-}$ in the Nambu -- Jona-Lasinio model}
\author{A. A. Osipov}
\email[]{aaosipov@jinr.ru} 
\altaffiliation{}
\affiliation{Bogoliubov Laboratory of Theoretical Physics, Joint Institute for Nuclear Research, Dubna, 141980, Russia}

\author{A. A. Pivovarov}
\email[]{tex\_k@mail.ru}
\thanks{}
\affiliation{Bogoliubov Laboratory of Theoretical Physics, Joint Institute for Nuclear Research, Dubna, 141980, Russia}

\author{M. K. Volkov}
\email[]{volkov@theor.jinr.ru} 
\thanks{}
\affiliation{Bogoliubov Laboratory of Theoretical Physics, Joint Institute for Nuclear Research, Dubna, 141980, Russia}

\begin{abstract}
The anomalous decays $f_1(1285)\to\rho^0\pi^+\pi^-$ and $a_1(1260)\to\omega\pi^+\pi^-$ violating natural parity for vectors and axial-vectors are studied in the framework of the Nambu -- Jona-Lasinio model. We consider the Lagrangian with $U(2)_L\times U(2)_R$ chiral symmetric four quark interactions. The theory is bosonized and corresponding effective meson vertices are obtained in the leading order of $1/N_c$ and derivative expansions. The uncertainties related with the surface terms of anomalous quark triangle diagrams are fixed by the corresponding symmetry requirements. We make a numerical estimate of the decay widths $\Gamma (f_1(1285)\to\rho^0\pi^+\pi^-)=2.78\, \mbox{MeV}$ and $\Gamma (a_1(1260)\to\omega\pi^+\pi^-)=87\, \mbox{keV}$.  Our result on the $f_1(1285)\to\rho^0\pi^+\pi^-$ decay rate is in a good agreement with experiment. It is shown that a strong suppression of the $a_1(1260)\to\omega \pi\pi$ decay is a direct consequence of destructive interference between box and triangle anomalies.
\end{abstract}

\maketitle

\section{Introduction}
The QCD perturbation theory is not applicable to the low-energy physics of hadrons ($E < 2\,\mbox{GeV}$). As a rule, in this region of energies one applies various phenomenological models based on an approximate chiral symmetry of strong interactions. The low-energy effective theory of QCD, known as the chiral perturbation theory \cite{Weinberg67,Weinberg79,Gasser83,Gasser84,Gasser85}, is one of such successful approaches at $E < m_\rho$. To extend the calculational scheme up to order ${\cal O}(p^6)$, it incorporates the lowest resonance spin-1 states implementing the appropriate QCD short-distance constraints \cite{Ecker89,Ecker89npb,Knecht01}. Another well-known approach is the famous Nambu -- Jona-Lasinio (NJL) model \cite{Nambu61a,Nambu61b} which incorporates the dynamical mechanism of spontaneous chiral symmetry breaking in hadron matter. Later on, this original idea of Nambu has been reinterpreted in terms of quarks and successfully implemented to the construction of the local effective meson Lagrangians including not only spin-0 states, but also the vector and axial-vector resonances \cite{Eguchi:1976iz,Volkov:1982zx,Ebert:1982pk,Volkov:1984kq,Volkov:1986zb,Ebert:1985kz,Vogl:1991qt,Klevansky:1992qe,Volkov:1993jw,Ebert:1994mf,Hatsuda:1994pi,Volkov:2006vq}. 

Our study here is based on the NJL model approach. The most successful results in this model are obtained for pseudoscalar and vector mesons. The description of scalars and axial-vectors is more problematic and still requires additional theoretical efforts. The recent progress here is related with the study of the anomalous radiative decays of the axial-vector $f_1(1285)$ and $a_1(1260)$ mesons \cite{Osipov:2017ray}. These vertices belong to the AVV-type and have several restrictions from the QCD low-energy theorems: the Adler-Bardeen theorem \cite{Adler69}, the Landau-Yang theorem \cite{Landau48,Yang50}, and so on. Presently, there is a growing interest in their theoretical and experimental investigations. This includes a recent measurement of the branching fraction of the $\tau\to f_1\pi\nu_\tau$ decay \cite{Lees:2012ks} and its theoretical description given in the different approaches \cite{Li:1996md,Calderon:2012zi,Vishneva:2014lla,Volkov18,Oset18}. In a number of works the electromagnetic vertices $f_1\rho\gamma$ and $f_1\omega\gamma$ have been considered \cite{He17,Kochelev09,Harvey09}. The vertex $f_1\gamma\gamma$ is important in the study of the hyperfine structure of muonic hydrogen \cite{Kochelev17}. There are also predictions for the $f_1\gamma$ and $a_1\gamma$ production from the $e^+e^-$ primary beams \cite{Volkov:2017qag}. All these studies should clarify the nature of $f_1(1285)$ and $a_1(1260)$ mesons.         

In this work, we calculate the $f_{1}(1285)\to \rho^{0}\pi^{+} \pi^{-}$ and $a_{1}(1260)\to\omega\pi^{+}\pi^{-}$ decay widths assuming the $q\bar q$ nature of $f_1(1285)$ and $a_1(1260)$. The first process has been considered in \cite{Gomm84} in the massive Yang-Mills approach, and in \cite{Meissner90} in the generalized hidden symmetry framework. In both cases the decay channel $f_1\to\rho^0\rho^0\to \rho^0\pi^+\pi^-$ has been neglected. We take into account this mode here. One of the purposes is to test the structure of an effective $AVV$-vertex, obtained recently in \cite{Osipov:2017ray}, in the case when one of the particles is off mass shell. The other goal is to study the structure of the box anomalous diagrams. Their contribution interferes with the triangle anomalies. We show that the result of this interference is controlled by the QED Ward identities through a mechanism of the vector meson dominance (VMD). The latter issue has been also addressed in \cite{Osipov18prd}. We are not aware of works where the decay width of $a_{1}(1260) \to \omega\pi^{+}\pi^{-}$ has been obtained. So, we hope that our study of this mode is helpful as a benchmark for future measurements.   

In our calculations we use the effective meson Lagrangian derived by bosonization of the NJL quark model \cite{Volkov:1986zb,Ebert:1985kz}. The local vertices of this Lagrangian arise in the long wavelength regime through a gradient expansion of the one-loop quark diagrams \cite{Kikkawa76}. The coefficients of the gradient expansion [masses and coupling constants] are expressed in terms of the model parameters, i.e. are known. That is essentially different from \cite{Meissner90} where amplitudes of $f_1(1285)\to\rho^+\pi^-\pi^0$ and $a_{1}(1260) \to \omega\pi^{+}\pi^{-}$ decays have been expressed in terms of unknown coupling constants. To make a progress here one should calculate these couplings. The NJL model gives such a possibility. We restrict our consideration to the tree-level Feynman diagrams. It makes conclusions to be valid to lowest order in $1/N_c$, where $N_c$ is the number of colors in QCD. 

Let us remind that resonances are narrow for large $N_c$, with widths of order $1/N_c$ \cite{Hooft74a,Hooft74b,Witten79}. This implies that one should neglect widths in the tree-level amplitudes unless resonances reach their on-shell peaks in the physical region. Since the zero-width propagators do not lead to the one-particle poles when one integrates over phase-space of the $f_{1}(1285)\to \rho^{0}\pi^{+} \pi^{-}$ and $a_{1}(1260)\to\omega\pi^{+}\pi^{-}$ decays, we may stay at leading $1/N_c$ order, and work in the zero-width approximation. This approximation can be improved by considering the next to leading order corrections in $1/N_c$. Such a step certainly would allow to take into account the finite widths of resonances. However, it would also require to take into account the one-loop meson diagrams. This seems too complicated for an initial study of these processes. That is why this issue will not be addressed here. 

Can $f_1(1285)\to\rho^0\pi^+\pi^-$ decay be successfully described in leading order of $1/N_c$ expansion? 	
It is quite plausible that this is possible. The reasoning is that a tree-level approximation has been already used in \cite{Gomm84} and \cite{Meissner90}. The derivation in \cite{Gomm84} led to a rather low value for the decay width $\Gamma (f_1(1285)\to\rho^0\pi^+\pi^-)\simeq 1$ to 1.167 MeV [for $m_{a_1}=1275$ MeV, and $m_{a_1}=1200$ MeV correspondingly] compared to the experimental value quoted by the Particle Data Group (PDG) \cite{Agashe:2014kda}
\begin{equation}
\label{expf1}
\Gamma^{exp} (f_{1}(1285) \to \rho^{0} \pi^{+} \pi^{-})= 2.651^{+ 0.169}_{-0.145} \textrm{ MeV}.
\end{equation}
However, this value is not actually a leading order $1/N_c$ result [the authors took into account a finite width of the $a_1(1260)$ meson, which is a non-leading contribution]. The zero-width calculations made in \cite{Meissner90} showed that $1/N_c$ expansion can be relevant to the question.

Despite the obvious similarity of $f_1(1285)\to \pi^+\pi^-\gamma$ and $f_1(1285)\to\rho^0 \pi^+\pi^-$ decays, the  role of the intermediate vector $\rho(770)$ and axial-vector $a_1(1260)$ states here is different. In the radiative decay, the contribution of the $\rho(770)$ exchange is dominated by the real pole in the physical region \cite{Osipov18prd}. On the contrary, a kinematic region for the process $f_1(1285)\to\rho^0 \pi^+\pi^-$  is such that a tree-level amplitude has no one-particle pole. It makes the $\rho$-exchange contribution to be rather small. On the contrary, a nearby on-shell singularity of the $a_1$ propagator enhances the $a_1$-exchange channel.

The paper is organized as follows. In Sec.~\ref{NJL-lagran} we review briefly the NJL model and establish our notations. In Sec.~\ref{f1} we derive the decay width of $f_1(1285)\to\rho^0 \pi^+\pi^-$. The different channels are analyzed in detail. Sec.~\ref{a1} is devoted to the $a_1(1260)\to\omega \pi^+\pi^-$ mode. We follow here the same strategy as for the $f_1(1285)$ decay.  We end with a short summary and conclusions in
Sec.~\ref{concl}. The moral seems to be that to describe decay widths of the processes considered one can use the leading order of the large $N_c$ expansion. However, if one wants to obtain the detailed information about other characteristics of $a_1(1260)$ and $f_1(1285)$ mesons one should go beyond the leading order result. 

\section{The Lagrangian of the NJL model}
\label{NJL-lagran}
Let us consider the extended NJL model with the $U(2)_L\times U(2)_R$ chiral symmetric four quark interactions \cite{Ebert:1982pk}. The Lagrangian density 
\begin{eqnarray}
\label{lag}
&&{\cal L}=\bar q(i\gamma^\mu\partial_\mu -{\cal M})q + {\cal L}_{S} +{\cal L}_{V}, \\
\label{lagsp}
&&{\cal L}_{S}=\frac{G_S}{2}\left[(\bar qq)^2+(\bar qi\gamma_5\vec\tau q)^2 \right], \\
\label{lagva}
&&{\cal L}_{V}=-\frac{G_V}{2}\left[(\bar q\gamma^\mu\vec\tau q)^2+(\bar q\gamma^\mu\gamma_5\vec\tau q)^2 \right]
\end{eqnarray}
includes spin-0, $G_S$, and spin-1, $G_V$, four-quark couplings; ${\cal M}=\hat m\tau_0$, $\hat m = \hat m_u=\hat m_d$ are the current quark masses (the isospin symmetry is assumed); $\tau_0$ is a unit $2\times 2$ matrix, $\vec\tau$ are the $SU(2)$ Pauli matrices; $\gamma^\mu$ are the standard Dirac matrices in four dimensional Minkowski space; in the notation of the quark field $q$ the color, isospin and 4-spinor indices are suppressed.

After introducing bosonic variables in the corresponding generating functional one obtains the equivalent bi-linearised form of multi-quark interactions, i.e., the Yukawa-type vertices describing the couplings of the collective meson fields with the quark-antiquark pairs. For our purpose here we need only the following part of the Lagrangian density
\begin{eqnarray}
\label{Lagrangiane}
\Delta {\cal L}_{int}&=&\frac{g_\rho}{2}\bar q\gamma^{\mu}\left[\gamma_{5}\left(f_{1\mu}+\vec\tau\vec{a}'_{1\mu}\right)+\left( \omega_{\mu} +\vec\tau\vec{\rho}_{\mu}\right) \right]q \nonumber \\
&+&ig_{\pi}\bar q\gamma_{5}\vec\tau\vec\pi q.
\end{eqnarray}
Here $q$ is the constituent quark field with up and down flavours; the $\vec\pi$, $\vec{\rho}_\mu$ and $\omega_\mu$ are the field operators associated with the iso-triplet of pions $\pi(140)$, vector $\rho (770)$ and $\omega (782)$-mesons; $f_{1\mu}$ describes the iso-singlet axial-vector $f_1(1285)$-meson (for simplicity we take $f_1(1285)$ to be the ideally mixed combination, corresponding to its status as an axial $\omega$), and $\vec{a}'_{1\mu}$ stands for the unphysical axial-vector fields that should be redefined to avoid the $\vec\pi -\vec{a}'_{1\mu}$ mixing. 

Since the free part of the meson Lagrangian following from evaluation of the one-quark-loop self-energy diagrams must preserve its canonical form, one should renormalize the bare meson fields by introducing the Yukawa coupling constants $g_\pi$ and $g_\rho$ in Eq. (\ref{Lagrangiane}). To absorb infinities of self-energy graphs, these couplings depend on the divergent integral $I_2$ which is regularized in a standard way \cite{Volkov:1986zb}
\begin{equation}
\label{g}
g_{\rho} = \sqrt{\frac{3}{2I_2}}, \quad g_{\pi} = \sqrt{\frac{Z}{4I_2}},
\end{equation}
where 
\begin{eqnarray}
I_{2} &=& -i\frac{N_{c}}{(2\pi)^{4}}\!\int\!  \mathrm{d}^{4}k \, \frac{\theta(\Lambda^{2} + k^2)}{(m^{2} - k^2)^{2}} \nonumber \\ 
&=&\frac{N_{c}}{(4\pi)^{2}}\left[\ln\left(1+\frac{\Lambda^2}{m^2}\right)-\frac{\Lambda^2}{\Lambda^2+m^2}\right].
\end{eqnarray}
As usual, we assume that the quantum corrections are valid only when the relevant momenta are less than the cut-off momentum $\Lambda$, which also has the meaning of the characteristic scale of spontaneous chiral symmetry breaking, defining through the gap equation 
\begin{equation}
\label{gap}
m-\hat m= mG_S I_1,
\end{equation}
where
\begin{equation}
I_1=\frac{N_c}{2\pi^2}\left[\Lambda^2-m^2\ln\left(1+\frac{\Lambda^2}{m^2}\right)\right],
\end{equation}
the masses $m$ of constituent quarks $q$. It is assumed that the strength of the quark interactions is large enough, $G_S>2\pi^2/(N_c\Lambda^2)$, to generate a non-trivial, $m\neq 0$, solution of Eq. (\ref{gap}) [even if the current quarks would be massless]. The non-zero value of $m$ is held to signal the condensation of quark-antiquark pairs in the vacuum, i.e. dynamical chiral symmetry breaking. 

The parameter $Z$ in (\ref{g}) appears as a result of elimination of the $\vec\pi -\vec a'_1$ transitions. For that one should redefine the axial-vector field
\begin{equation}
\label{pa-trans}
\vec a'_{1\mu}=\vec a_{1\mu}+\sqrt{\frac{2Z}{3}}\kappa m\partial_\mu\vec\pi,
\end{equation}   
where $\vec a_{1\mu}$ represents a physical state $a_1(1260)$. A dimensional parameter $\kappa$, related with $Z$ by $1-2\kappa m^2=Z^{-1}$, should be fixed by requiring that the meson Lagrangian does not contain the $\vec\pi-\vec a_{1\mu}$ mixing. It gives 
\begin{equation}
\label{kappa}
\frac{1}{2\kappa}=m^2+\frac{1}{16 G_V I_2}=m^2+\frac{m_\rho^2}{6}=\frac{m_{a_1}^2}{6},
\end{equation}
where the last two equalities are a consequence of the mass formulas of the model. 

The model has four parameters: $G_S$, $G_V$, $\hat m$, and $\Lambda$. To fix them we use the following empirical data. From the $\rho\to\pi\pi$ decay width we find that $\alpha_\rho =g_\rho^2/(4\pi )=3$. It gives $I_2=1/(8\pi )$, and $\Lambda/m=4.48$. Using the mass of the $\rho$ meson as a second input value, $m_\rho =775\,\mbox{MeV}$, we find $G_V$ from the mass formula of the $\rho$ meson
\begin{equation}
G_V=\frac{3}{8m_\rho^2I_2}=\frac{3\pi}{m_\rho^2}=1.57\times 10^{-5}\,\mbox{MeV}^{-2}.
\end{equation}
The coupling constant $g_\pi$ fulfilles at the quark level the celebrated Goldberger-Treiman relation $g_\pi =m/f_\pi$, where $f_\pi =93\, \mbox{MeV}$ is a coupling of the $\pi^-\to\mu^-\bar\nu_\mu$ weak decay which we use as a third input. Then Eq. (\ref{g}) gives 
\begin{equation}
\label{f}
6m^2= Z g_\rho^2 f_\pi^2.
\end{equation}
Using (\ref{kappa}) this equation can be transformed to the formula 
\begin{equation}
\label{KS}
m_\rho^2=\left(\frac{Z}{Z-1}\right) g_\rho^2f_\pi^2
\end{equation}  
that gives $Z=2.188$, or $2\kappa m^2=0.543$. In this case, from the relation (\ref{f}) one finds $m=344.8\, \mbox{MeV}$, and $\Lambda =4.48 m=1544.7\,\mbox{MeV}$.

Taking as the final input the value of the pion mass, $m_{\pi} =138\,\mbox{MeV}$, we are left with the system of two equations, Eq. (\ref{gap}) and a pion mass formula 
\begin{equation}
m_{\pi}^2 =\frac{\hat m g_\pi^2}{mG_S}=\frac{\hat mm}{G_Sf_\pi^2}, 
\end{equation}
to find the values of the current quark mass $\hat m$, and the coupling $G_S$. Solving this system, we obtain 
\begin{eqnarray}
&& G_S=\frac{\displaystyle m^2}{\displaystyle m_\pi^2f_\pi^2+m^2I_1} =3.24 \times 10^{-6}\,\mbox{MeV}, \\
&& \hat m=m\left(1-G_SI_1 \right)=1.55\,\mbox{MeV}.
\end{eqnarray}

It follows then that the mass of the $a_1$ meson is given by $m_{a_1}=\sqrt{Z}m_\rho =1146 \,\mbox{MeV}$. This result agrees well with the Weinberg's prediction $m_{a_1}=\sqrt{2}m_\rho =1096 \,\mbox{MeV}$ \cite{Weinberg67a} made on the basis of spectral-function sum rules, which are valid in QCD for $m_\pi =0$, and KSRF formula \cite{Kawarabayashi66,Riazuddin66} for the $\rho$ coupling to the isospin current [in our case there is a similar relation (\ref{KS})]. It also agrees with a theoretical analysis of \cite{Pich10}, where the excellent agreement with our present experimental knowledge of $\tau\to\pi\pi\pi\nu_\tau$ spectrum and branching ratio \cite{Barate98} has been obtained and the characteristics of $a_1$ meson have been carefully extracted, giving $m_{a_1}=1120\,\mbox{MeV}$, and $\Gamma_{a_1}=483\,\mbox{MeV}$. On the other hand, our result is a little low compared to the value $m_{a_1}=1230\pm 40\,\mbox{MeV}$ quoted by the Particle Data Group \cite{Agashe:2014kda}. About the larger value of the $a_1(1260)$ mass has been recently reported by the COMPASS collaboration: $m_{a_1}=1298^{+13}_{-22}\,\mbox{MeV/c}^2$ with $\Gamma_{a_1}=400^{+0}_{-100}\,\mbox{MeV/c}^2 $ \cite{Wallner17}. Notice that their data are accumulated from the study of the channel $p + \pi^-\to \pi^-\pi^-\pi^+ + p_{ recoil}$, for which COMPASS has acquired the so far world's largest dataset of roughly 50M exclusive events using an $190\,\mbox{GeV/c}$ $\pi^-$ beam. 
 
In the following, the necessary effective meson vertices [together with the corresponding coupling constants] will be obtained from (\ref{Lagrangiane}) by calculating the one-quark-loop diagrams and taking out from them only the leading terms in the derivative expansion which dominate in the long-wavelength approximation. The decay amplitudes are given by a sum of tree-level diagrams involving the exchange of physical mesons. This approach is consistent with a picture arising in the large $N_c$ limit of QCD \cite{Hooft74a,Hooft74b,Witten79}.

\section{The process $f_{1}(1285) \to \rho^{0} \pi^{+} \pi^{-}$}
\label{f1}
The partial width for the observed decay mode of the axial-vector meson $f_1(1285)\to\rho^0\pi^+\pi^-$ can be estimated in the NJL model by considering the following tree-level contributions: (a) the vector $\rho^0$-meson exchange channel $f_1\to\rho^0\rho^0\to \rho^0\pi^+\pi^-$; (b) the axial-vector $a_1^\pm$-meson  exchange channel $f_1\to\pi^{\pm} a_1^{\mp}\to \pi^{\pm}\pi^{\mp}\rho^0$; (c) the direct decay which is described by the quark box diagram. 
\subsection{Kinematic invariants, the physical region and a structure of the amplitude}
\label{ss3.1}
In the discussion of the decay $f_{1}(l) \to \rho^{0}(p) + \pi^{+}(p_+) + \pi^{-}(p_-)$ we will use the standard invariant quantities which can be constructed from 4-momenta of particles $l, p, p_+$ and $p_-$, namely  
\begin{eqnarray}
s&=&(l-p)^2=(p_++p_-)^2, \nonumber \\
t&=&(l-p_+)^2=(p+p_-)^2, \nonumber \\
u&=&(l-p_-)^2=(p+p_+)^2.
\end{eqnarray}
Only two of them are independent variables, because of the relation $s+t+u=h$, where $h=m_f^2+m_\rho^2+2m_\pi^2$. From the law of conservation of 4-momentum one finds the intervals for physical values of these variables
\begin{eqnarray}
&&4m_\pi^2\leq s \leq (m_f-m_\rho )^2 , \nonumber \\
&&(m_\rho +m_\pi )^2 \leq t, u \leq (m_f -m_\pi )^2.
\end{eqnarray}
The equation 
\begin{equation}
\label{bound}
t^2-t(h-s)+\frac{1}{4}\left[(h-s)^2-D(s) \right]=0,
\end{equation}
where
\begin{eqnarray}
&&D(s)=\frac{1}{s}(s-4m_\pi^2)\lambda (s, m_f^2, m_\rho^2), \\
&& \lambda (x, y, z)= (x-y-z)^2 - 4 y z \nonumber \\
&& = [x-(\sqrt{y}+\sqrt{z})^2][x-(\sqrt{y}-\sqrt{z})^2],
\end{eqnarray}
defines a curve which is the boundary of the physical region for the decay channel. There are two positive values of $t$ for each value of $s$. These two roots of the quadratic Eq. (\ref{bound}) are the endpoints of the closed interval for physically permissible values of $t_-\leq t\leq t_+$
\begin{equation}
t_\pm =\frac{1}{2} \left(h-s \pm \sqrt{ D(s)} \right).
\end{equation}
Notice, that $D(4m_\pi^2)=D((m_f-m_\rho )^2)=0$.

One can see that the $\rho^0$-resonance exchange contribution has no pole at physical values of meson masses: $m_\pi =138\,\mbox{MeV}$, $m_\rho =775\,\mbox{MeV}$, $m_{f_1}=1282\,\mbox{MeV}$. Indeed, the one-particle pole in the amplitude, if it is, comes out of the factor $(m_\rho^2-s)^{-1}$. However, the physical values of $s$ belong to the interval $2m_\pi \le \sqrt{s}\le m_{f_1}-m_\rho$, or numerically $276\,\mbox{MeV}\leq \sqrt{s}\leq 507\,\mbox{MeV}$, which is quite distant from the $\rho$-meson mass.

The $a_1$ meson exchange amplitudes include one of the factors $(m_{a_1}^2-t)^{-1}$, or $(m_{a_1}^2-u)^{-1}$. The physical region of the kinematic variables $t$ and $u$ is such that $913\,\mbox{MeV}\leq \sqrt{t}, \sqrt{u}\leq 1144\,\mbox{MeV}$. This certainly indicates that although there is no real pole here the contribution is sensitive to the mass of the $a_1$ meson. In particular, this channel will dominate if the mass of the $a_1$ is about the model estimate $m_{a_1}=1146\,\mbox{MeV}$. On the contrary, at large values of $m_{a_1}=1230-1290\,\mbox{MeV}$ the $a_1$ exchange may lead approximately to the same order contribution as the $\rho$ exchange. This reasoning show that the decay mode $f_1\to\rho^0\pi^+\pi^-$ may supply us with interesting information on the $a_1$-meson characteristics.  

The amplitude of the process [as it follows from the NJL model calculations below] may be parametrized as
\begin{eqnarray}
\label{amplitude}
T&=& ie_{\mu\nu\alpha\beta} \epsilon^\beta (l)\epsilon^*_\gamma (p)\left[ g^{\alpha\gamma}\left( F_1l^\mu p^\nu_+ +F_2l^\mu p^\nu_- + F_3 p_+^\mu p_-^\nu \right)\right.  \nonumber\\
&+& \left. F_4 p^\alpha l^\gamma p_+^\mu p_-^\nu \right],
\end{eqnarray}
where $\epsilon_\beta (l)$,  $\epsilon_\gamma (p)$ are the polarization vectors of the $f_1$ and $\rho$ mesons. In the following we will obtain the explicit expressions for the form factors $F_a$, $a=1,2,3,4$ in the framework of the NJL model at leading order of  $1/N_c$ and derivative expansions. The different channels contribute to the sum independently 
\begin{equation}
F_a = F_a^{(\rho )}+F_a^{(a_1)}+F_a^{(d)}.
\end{equation}
Here, $F_a^{(\rho )}$ is a contribution of the $\rho^0$-exchange channel (a), $F_a^{(a_1 )}$ describes the axial-vector $a_1^\pm$ exchange mode (b), and the direct interaction (c) is presented by the form factor $F_a^{(d)}$.  

\subsection{The $\rho^0(770)$ exchange channel}
\label{ss3.2}
The resonance exchange mode $f_1\to\rho^0\rho^0\to \rho^0\pi^+\pi^-$ in the NJL model can be described by the following Lagrangian densities. 

The anomalous $f_1\rho^0\rho^0$ vertex can be easily obtained from the $f_1\rho^0\gamma$ vertex \cite{Osipov:2017ray}. For that one should replace the electromagnetic field by the $\rho^0$ field, electric charge $e$ by the coupling $g_\rho$, and introduce the factor $1/2$ accounting for identity of two $\rho^0$-meson states in the Lagrangian. As a result we obtain
\begin{equation}
\label{frr}
{\cal L}_{f_1\rho^0\rho^0}=\frac{g_\rho^3N_c}{3(8\pi m)^2} e^{\mu\nu\alpha\beta} \rho^0_{\mu\nu} \left(\rho^0_{\sigma\alpha} \stackrel{\leftrightarrow}{\partial^\sigma}\! \! f_{1\beta}\right),
\end{equation}
where $(a\stackrel{\leftrightarrow}{\partial_\mu} b)=a\partial_\mu b-(\partial_\mu a) b$, and $\rho_{\mu\nu}$ stands for the field strength $\rho_{\mu\nu}=\partial_\mu\rho_\nu -\partial_\nu\rho_\mu$. 

\begin{figure}
\includegraphics[width=0.45\textwidth]{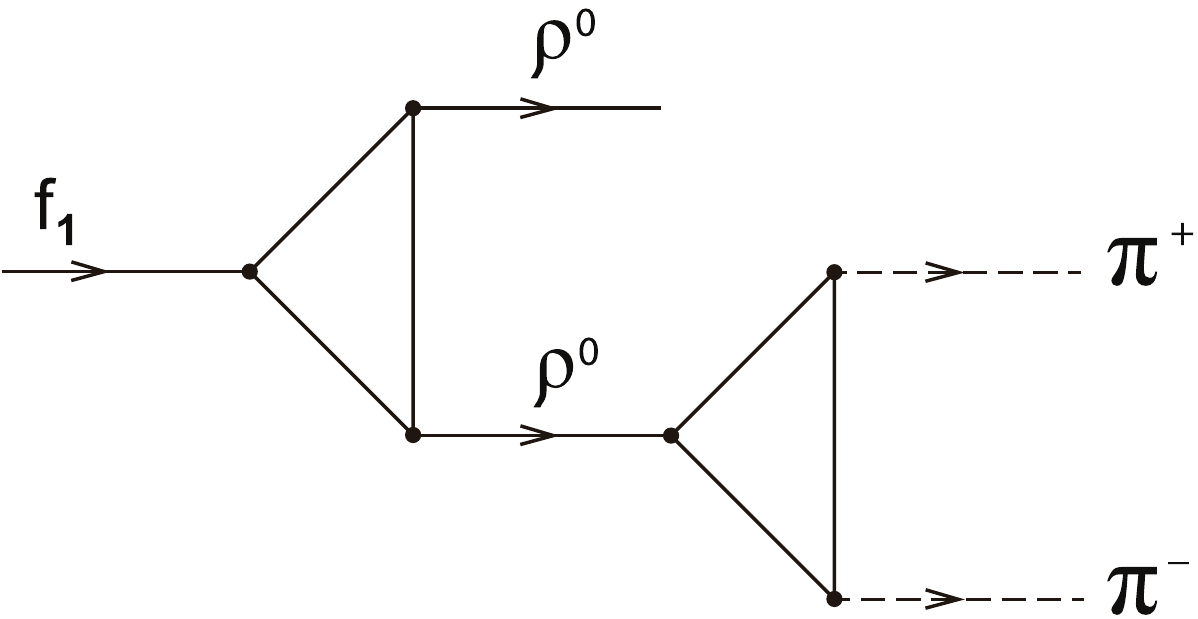}
\caption{The Feynman diagrams describing the $\rho^0$ exchange mode for the $f_{1}(1285) \to \rho^{0} \pi^{+} \pi^{-}$ decay. It is assumed [for all figures in the text], that each pion line represents the sum of two types of couplings of the pion with the quark-antiquark pair: the pseudoscalar one $\bar q\gamma_5\pi q$ and the axial-vector one $\bar q \gamma^\mu\gamma_5\partial_\mu\pi q$.}
\label{fig1}      
\end{figure}

Notice, that  the effective vertex ${\cal L}_{f_1\rho^0\rho^0}$ is given by the next to the leading order term in the derivative expansion of the anomalous quark triangle diagram $f_1\rho^0\rho^0$ shown in Fig.~\ref{fig1}. Actually, one would expect here the contribution linear in momenta. Bose symmetry requires that it would have a form 
\begin{equation}
\label{rem}
{\cal L}'_{f_1\rho^0\rho^0} \propto e^{\alpha\beta\mu\nu}f_{1\alpha }\rho^0_\mu \partial_\beta \rho^0_\nu .
\end{equation} 
This form, however, is not compatible with the idea of vector dominance. In the real world with electromagnetic interactions included, this vertex would generate the gauge symmetry breaking contributions to the $f_1\to\rho^0\gamma$ and $f_1\to\gamma\gamma$ amplitudes. So, in fact, (\ref{rem}) is not consistent with the QED Ward identities.  Let us also notice, that a superficial linear divergence appears in the course of evaluation of the overall finite $f_1\rho^0\rho^0$ triangle integral. Shifts in the internal momentum variable of the closed quark loop integrals induce an arbitrary finite surface term contribution of the type (\ref{rem}). Thus, one can always choose the free coupling of the surface term to vanish (\ref{rem}). This avoids contradiction with Ward identities.    

The nonanomalous $\rho\pi\pi$ vertex in Fig.~\ref{fig1} is described by the Lagrangian density  
\begin{eqnarray}
\label{rpp}
{\cal L}_{\rho^0\pi^+\pi^-}&=&-ig_\rho \rho^0_\mu (\pi^+\! \!\stackrel{\leftrightarrow}{\partial^\mu}\! \pi^- ) \nonumber \\ 
&+&ig_\rho \frac{Z-1}{m_{a_1}^2}\,\rho^0_{\mu\nu} \partial^\mu\pi^+\partial^\nu\pi^- ,
\end{eqnarray}   
where, in the following, we neglect the second term in (\ref{rpp}). The reasoning for this is that it has a small factor $s(Z-1)/(2m_{a_1}^2)=s\kappa m^2/m_\rho^2$ [compared with the factor 1 of the first term], which varies from 0.03 to 0.1 in the kinematic region of $s$.  

With the use of these Lagrangian densities we find the $\rho^0$ exchange contribution to the amplitude of the process shown in Fig.~\ref{fig1}. The result is 
\begin{eqnarray}
\label{rhoexch}
F_1^{(\rho )}&=& \left(\frac{\alpha_\rho^2}{2m^2}\right)\frac{m_{f_1}^2+m_\rho^2 -2m_{f_1}(\varepsilon -\varepsilon_- )}{m_\rho^2 - s  },  \nonumber \\
F_2^{(\rho )}&=&\left(\frac{-\alpha_\rho^2}{2m^2}\right)\frac{m_{f_1}^2+m_\rho^2-2m_{f_1}(\varepsilon-\varepsilon_+ )}{m_\rho^2 - s  }, \nonumber \\
F_3^{(\rho )}&=&\left(\frac{2\alpha_\rho^2}{m^2}\right)\frac{m_{f_1}^2+m_\rho^2-m_{f_1}\varepsilon}{m_\rho^2 - s  }, \nonumber \\
F_4^{(\rho )}&=&\left(\frac{-\alpha_\rho^2}{m^2}\right)\frac{1}{m_\rho^2 - s  }, 
\end{eqnarray}
where $\varepsilon, \varepsilon_\pm$ are the energies of the rho meson and charged pions in the rest frame of the $f_1(1285)$-meson. 

Notice that this channel [through the $\rho^0\to\gamma$ transition] gives the determining contribution to the decay width of $f_1(1285)\to \pi^+\pi^-\gamma$ \cite{Osipov18prd}. Conversely, the diagram shown in Fig.~\ref{fig1} is not so important for the $f_1(1285)\to \rho^0\pi^+\pi^-$ decay. Indeed, its contribution to the decay width is $\Gamma (f_1(1285)\to \rho^0 \pi^+\pi^-)=37\,\mbox{keV}$. We conclude that this channel is strongly suppressed in comparison with $a_1$ exchange channel [as it will be shown in Sec.~\ref{ss3.3}], but it is still worth to be taken into account due to their constructive interference.

\subsection{The $a_1(1260)$ exchange channel}
\label{ss3.3}
\begin{figure}
\resizebox{0.40\textwidth}{!}{%
 \includegraphics{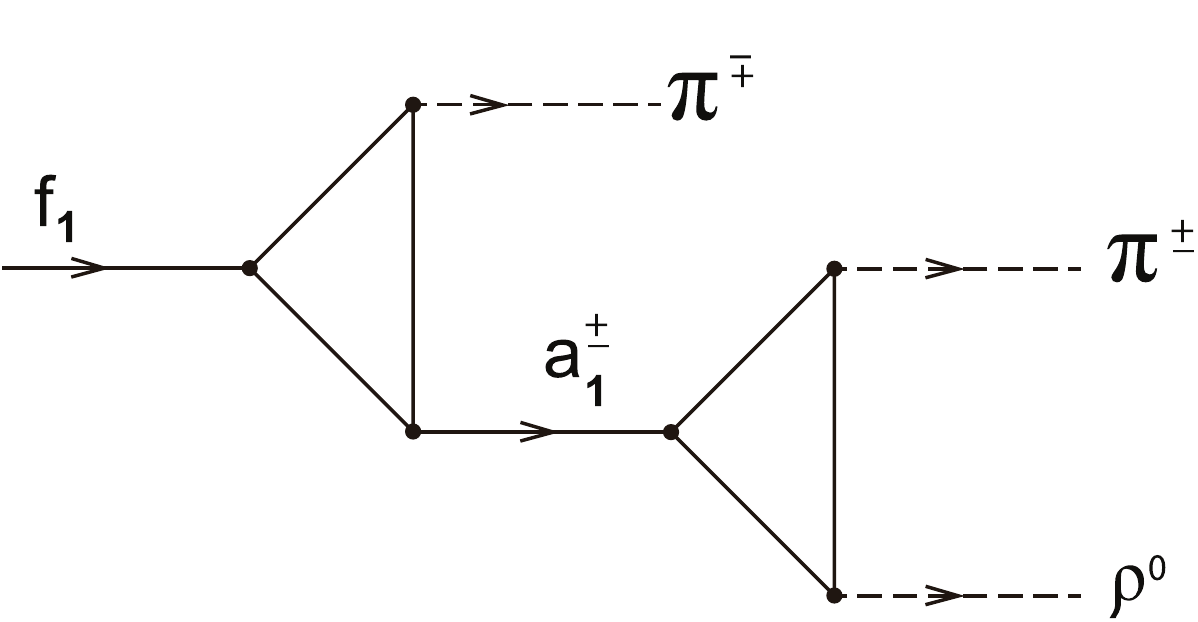}}
\caption{Two Feynman diagrams describing the $a_1^+$ and $a_1^-$ exchange modes for the $f_{1}(1285) \to \rho^{0} \pi^{+} \pi^{-}$ decay.}
\label{fig2}      
\end{figure}

To describe the $a_1$ exchange modes $f_1\to \pi^\pm a_1^\mp \to \pi^+\pi^-\rho^0$, shown in Fig.~\ref{fig2}, we use the nonanomalous Lagrangian density \cite{Osipov17ap}
\begin{eqnarray}
\label{apr}
{\cal L}_{a_1\pi\rho^0}&=&if_\pi g_\rho^2  Z \left[ \rho_\mu^0 a_1^{-\mu} \pi^+ \right. \nonumber \\
 &\!\!\!\!\!\!\!\!\!\!\!\!\!\!\!\! +&\!\!\!\!\!\!\!\!\!\! \left.\frac{1}{m_{a_1}^2}\left(a^-_{1\mu\nu}\rho^{0\mu} - a_{1}^{-\mu}\rho_{\mu\nu}^0\right)\partial^\nu \pi^+ \right]+h.c.,
\end{eqnarray}   
and the vertex which describes the anomalous $f_1a_1\pi$ interaction
\begin{equation}
\label{fap}
{\cal L}_{f_1a_1\pi}=g_{a} e^{\alpha\beta\mu\nu} f_{1\alpha} \partial_\mu\vec a_{1\beta} \partial_\nu\vec\pi ,
\end{equation} 
where 
\begin{equation}
\label{ga}
g_{a}=\frac{\alpha_\rho}{2\pi f_\pi}\left[1+(1-3a)\kappa m^2\right].
\end{equation}
The second term in the square brackets is due to the replacement (\ref{pa-trans}). The derivative coupling  $\bar q \gamma^\mu \gamma_5\partial_\mu\vec \pi \vec\tau q$ makes the corresponding triangle quark diagram linearly divergent, although the result of its evaluation is finite. As a consequence of this superficial divergence, an arbitrary finite surface term contribution proportional to $(1-3a)$ appears. Here $a$ is a dimensionless constant, controlling the magnitude of an arbitrary local part \cite{Jackiw00,Hiller01}. 

A corresponding contribution to the amplitude (\ref{amplitude}) is given by 
\begin{eqnarray}
\label{a1exch}
&&T^{(a_1)}=-ig_a \left(\frac{2\kappa m^2}{f_\pi}\right) e_{\mu\nu\alpha\beta} \epsilon^\beta (l) \epsilon^{*\alpha} (p) l^\mu p_+^\nu \nonumber \\
&&\left[1 + \frac{p^2}{m_{a_1}^2-t }\right]-(p_+\leftrightarrow p_-).
\end{eqnarray}
One can see that the contact part of this amplitude [the first term in the square brackets] would violate the gauge invariance, if one, following the idea of vector-meson dominance, switches to the related electromagnetic process [notice, that the second term in the square brackets does not contribute to the radiative decay $f_1(1285)\to\pi^+\pi^-\gamma$, because $p^2=0$ for a real photon]. Indeed, introducing the 4-vector $q^\nu=(p_+ -p_-)^\nu$, and using the four-momentum conservation law $l=p+p_++p_-$, one obtains 
$$
e_{\mu\nu\alpha\beta} l^\mu q^\nu =e_{\mu\nu\alpha\beta}\left( p^\mu q^\nu -2p_+^\mu p_-^\nu\right).
$$ 
If one replaces $\epsilon^*_\alpha (p)\to p_\alpha$ in (\ref{a1exch}), one finds that the term $\propto p_+^\mu p_-^\nu$ survives. This violates Ward identities. The point can be settled after considering the direct (box) part of the amplitude shown in Fig.~\ref{fig3} [see Sect.~\ref{ss3.4}]. 

To summarize, two diagrams with the $a_1$ exchange yield   
\begin{eqnarray}
F_1^{(a_1^-)}&=&-\frac{g_a}{f_\pi} \left(2\kappa m^2\right)\left[1+\frac{m_\rho^2}{m_{a_1}^2-t}\right],  \nonumber \\
F_2^{(a_1^+)}&=&\frac{g_a}{f_\pi} \left(2\kappa m^2\right)\left[1+\frac{m_\rho^2}{m_{a_1}^2-u}\right].
\end{eqnarray}

\subsection{The direct channel}
\label{ss3.4}
\begin{figure}
\resizebox{0.30\textwidth}{!}{%
\includegraphics{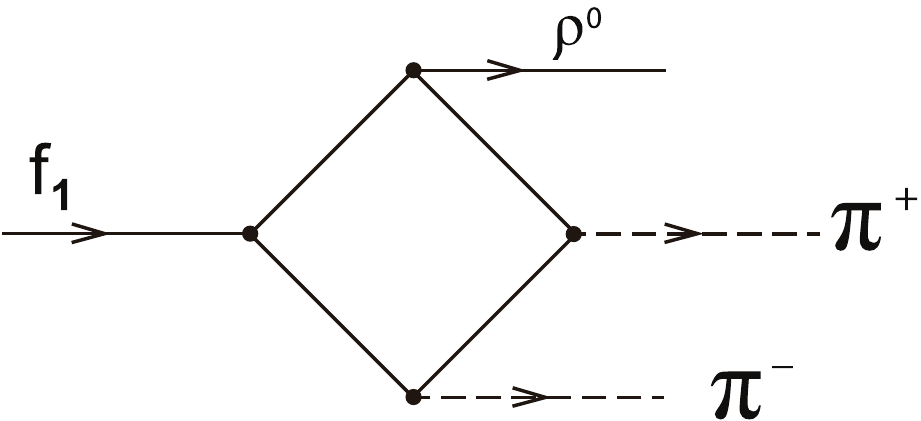}
}
\caption{The box Feynman diagrams for the $f_1 (1285) \to \rho^0\pi^{+} \pi^{-}$ decay. We do not show the diagrams which can be obtained by permuting the final states.}
\label{fig3}      
\end{figure}

Let us consider now the contribution to the decay amplitude $f_1(1285)\to \rho^0\pi^+\pi^-$ due to the quark box diagrams shown in Fig. \ref{fig3}. As usual, we will extract only the terms which are dominant at large distances, i.e. the local effective vertices with the smallest number of derivatives. This contribution contains information on the box AAAV anomaly. The calculations performed in a way explained above lead us to the amplitude  
\begin{eqnarray}
\label{boxcont}
T^{(d)}&=&i\frac{\alpha_\rho}{2\pi f_\pi^2} e^{\mu\nu\alpha\beta}\epsilon_\beta (l)\epsilon_\alpha^*(p)\left[(1-2\kappa m^2) p_\mu q_\nu \right. \nonumber \\
&-& \left. \kappa m^2 (4-\kappa m^2)p_+^\mu p_-^\nu \right].
\end{eqnarray} 
It can be easily seen that if we again resort to the radiative decay $f_1\to\pi^+\pi^-\gamma$ amplitude the term $\propto p_+^\mu p_-^\nu$ will break the gauge symmetry. The most efficient way of dealing with the issue is to sum all contact contributions and fix the free parameter $a$ by requiring the vanishing of the $p_+^\mu p_-^\nu$ term. Combining contact terms of Eqs.~(\ref{a1exch}) and (\ref{boxcont}), we find
\begin{eqnarray}
\label{ct}
&& T^{(a_1)}_{cont}+T^{(d)}= T^{(c)}=\frac{i\alpha_\rho}{2\pi f_\pi^2} e_{\mu\nu\alpha\beta} \nonumber \\
&&\times \epsilon^\beta (l)\epsilon^{*\alpha} (p)\left( A_1 p^\mu q^\nu +A_2 p_+^\mu p_-^\nu \right),
\end{eqnarray}  
where
\begin{eqnarray}
&& A_1=1-2\kappa m^2-2\kappa m^2 \left[1+(1-3a)\kappa m^2\right], \nonumber \\
&& A_2=(\kappa m^2)^2(5-12a).
\end{eqnarray}
At $a=5/12$ one finds that $A_2=0$. This solves the problem. This gives for $A_1$ 
\begin{equation}
A_1=1-4\kappa m^2+\frac{1}{2}(\kappa m^2)^2=\frac{2-Z}{Z}+\frac{(Z-1)^2}{8Z^2} .
\end{equation} 

Thus, the contact terms contribute to the amplitude as
\begin{equation}
\label{contact}
\label{contf}
F_1^{(c)}=-F_2^{(c)}=\frac{1}{2}F_3^{(c)}=\frac{\alpha_\rho}{2\pi f_\pi^2}A_1.
\end{equation}        
Correspondingly, the diagrams plotted in Figs.~\ref{fig2}, \ref{fig3} give the following contributions to the pertinent form factors  
\begin{eqnarray}
\label{a1+d}
F_1^{(a_1)}+F_1^{(d)}&=&F_1^{(c)}-\left(\frac{\alpha_\rho^2}{2}\right)\frac{(4-\kappa m^2)}{m_{a_1}^2-t},  \nonumber \\
F_2^{(a_1)}+F_2^{(d)}&=&F_2^{(c)}+\left(\frac{\alpha_\rho^2}{2}\right)\frac{(4-\kappa m^2)}{m_{a_1}^2-u}, \nonumber \\
F_3^{(a_1)}+F_3^{(d)}&=&F_3^{(c)},
\end{eqnarray}
where the relation 
\begin{equation}
\label{gafixed}
g_{a=\frac{5}{12}}\frac{2\kappa m^2}{f_\pi}= \frac{\alpha_\rho^2}{2m_\rho^2}\left(4-\kappa m^2\right) 
\end{equation}
has been used.  

Before we will present the result of our calculations in full details, it is instructive to show here the dominant role of the $a_1$-exchange contribution. In fact, as it follows from Eq.~(\ref{contact}), the sum of contact contributions is negligible: $\Gamma^{(c)} (f_1(1285)\to\rho^0\pi^+\pi^-)=1.2\,\mbox{keV}$. This is a consequence of the strong cancellation between a contact term in (\ref{a1exch}) and the contribution of the box diagram (\ref{boxcont}). One of the reasons is the Ward identities which control the value of a surface term fixing $a=5/12$. The size of this effect is quite large. To understand how this works, let us compare the $a_1$-exchange (\ref{a1exch}), calculated with $a=5/12$, $\Gamma^{(a_1)} (f_1(1285)\to\rho^0\pi^+\pi^-)=3.87\,\mbox{MeV}$ with Eq. (\ref{a1+d}), which gives lower value $\Gamma^{(d+a_1)} (f_1(1285)\to\rho^0\pi^+\pi^-)=2.22\,\mbox{MeV}$. The difference between these two numbers is an effect of the box diagram, which is taken into account in the latter case.   

\subsection{The $f_1(1285)\to\rho^0\pi^+\pi^-$  decay width}
\label{ss3.5}
The rate of the three-body decay $f_1(1285)\to\rho^0\pi^+\pi^-$ can be obtained from the standard formula 
\begin{equation}
\label{dG}
d\Gamma =\frac{|T|^2}{24m_{f_1}(2\pi )^3}d\varepsilon d\varepsilon_+
\end{equation}
where 
\begin{equation}
|T|^2=\sum_{i\leq j} \mbox{Re} \left(F_i F_j^* \right) T_{ij},
\end{equation}
\begin{equation}
F_i=F_i^{(\rho )}+F_i^{(a_1)}+F_i^{(d)},
\end{equation}
and
\begin{eqnarray}
T_{11}&=&m_{f_1}^2 \left(2\vec p_+^{\  2}+\Delta \right), \nonumber \\
T_{22}&=&m_{f_1}^2 \left(2\vec p_-^{\  2}+\Delta \right), \nonumber \\
T_{33}&=&2[(p_+p_-)^2- m_{\pi}^4]+ (m_{f_1}^2+m_\rho^2)\Delta ,  \nonumber \\
T_{44}&=&m_{f_1}^4 \vec p^{\ 2}\Delta ,   \nonumber \\
T_{12}&=&2m_{f_1}^2 \left(2\vec p_+\vec p_- -\Delta \right),  \nonumber \\
T_{13}&=&4m_{f_1} [m_\pi^2\varepsilon_- - (p_+ p_- )\varepsilon_+ ]  -2m_{f_1}^2\Delta , \nonumber \\
T_{23}&=&-4m_{f_1} [m_\pi^2\varepsilon_+ - (p_+ p_- )\varepsilon_- ] +2m_{f_1}^2\Delta , \nonumber \\
T_{14}&=&-T_{24}=-2m_{f_1}^3\varepsilon\Delta , \nonumber \\
T_{34}&=&2m_{f_1}^2 (\varepsilon m_{f_1} -m_\rho^2)\Delta .
\end{eqnarray}
Notice that 
\begin{eqnarray} 
&&m_\rho^2\Delta = (\vec p_+ \times\vec p)^2=(\vec p_- \times\vec p)^2=(\vec p_+ \times\vec p_-)^2 \nonumber \\
&&=\vec p_+^{\;  2}\vec p^{\;  2}-(\vec p_+\vec p)^2.
\end{eqnarray} 
Here all kinematic variables are given in the rest frame of the $f_1$ meson. In this reference system the invariant variables are 
\begin{eqnarray}
&& s=m_{f_1}^2 +m_\rho^2-2m_{f_1}\varepsilon , \nonumber \\
&& t=m_{f_1}^2+m_\pi^2-2m_{f_1}\varepsilon_+ , \nonumber \\
&& u=m_{f_1}^2+m_\pi^2 -2m_{f_1}(m_{f_1}-\varepsilon -\varepsilon_+ ).
\end{eqnarray}  
Thus, the physical region for independent variables $\varepsilon$ and $\varepsilon_+$ is given by the inequalities 
\begin{eqnarray}
\label{intervals}
&& m_\rho\leq \varepsilon \leq \frac{1}{2m_{f_1}}\left(m_{f_1}^2+m_\rho^2-4m_\pi^2\right), \nonumber \\
&& \frac{m_{f_1}-\varepsilon -\sqrt{\Omega (\varepsilon )}}{2}\leq \varepsilon_+ \leq \frac{m_{f_1}-\varepsilon +\sqrt{\Omega (\varepsilon )}}{2}
\end{eqnarray}
where
\begin{equation}
\Omega (\varepsilon )=(\varepsilon^2-m_\rho^2)\left(1-\frac{4m_\pi^2}{m_{f_1}^2 +m_\rho^2-2m_{f_1}\varepsilon}\right).
\end{equation}

Integrating in (\ref{dG}) over energies taken in the given intervals (\ref{intervals}) we find that the decay width of the process $f_{1}(1285) \to \rho^{0} \pi^{+} \pi^{-}$ is
\begin{equation}
\label{f1dw}
\Gamma(f_{1}(1285) \to \rho^{0} \pi^{+} \pi^{-}) = 2.78 \textrm{ MeV}.
\end{equation}

Thus, the picture can be summarized as follows. The $a_1$-exchange gives the major contribution because it is enhanced by a nearby singularity of the $a_1$ propagator. The box diagram almost cancels the contact part of (\ref{a1exch}) reducing decay width on 46\%. The $\rho$-exchange (\ref{rhoexch}) is small but its interference with other channels increases the result from $\Gamma^{(d+a_1)}=2.22\,\mbox{MeV}$ to the final value (\ref{f1dw}). This value is obtained in the leading order of $1/N_c$ expansion and agrees well with empirical data (\ref{expf1}). 

\section{The process $a_{1}(1260) \to \omega \pi^{+} \pi^{-}$}
\label{a1}
The calculation of the decay amplitude $a_{1}(l) \to \omega (p) +\pi^{+} (p_+) + \pi^{-}(p_-)$, where $l, p, p_+, p_-$ are the 4-momenta of corresponding particles, can be carried out in a similar way as was being done for the $f_{1}(1285) \to \rho^{0} \pi^{+} \pi^{-}$ decay in Sec.~\ref{f1}. The amplitude accumulates contributions from three different processes: (a) the $\rho^0$ exchange channel $a_1\to\omega\rho^0\to\omega\pi^+\pi^-$; (b) the $\rho^\pm$ exchange $a_1\to\pi^{\pm} \rho^{\mp}\to\pi^+\pi^-\omega$; and (c) the direct decay mode $a_1\to\omega\pi^+\pi^-$. The kinematic variables and the physical region can be easily obtained from the expressions presented in Sec.~\ref{ss3.1} and Sec.~\ref{ss3.5}.

\begin{figure}
\resizebox{0.40\textwidth}{!}{%
 \includegraphics{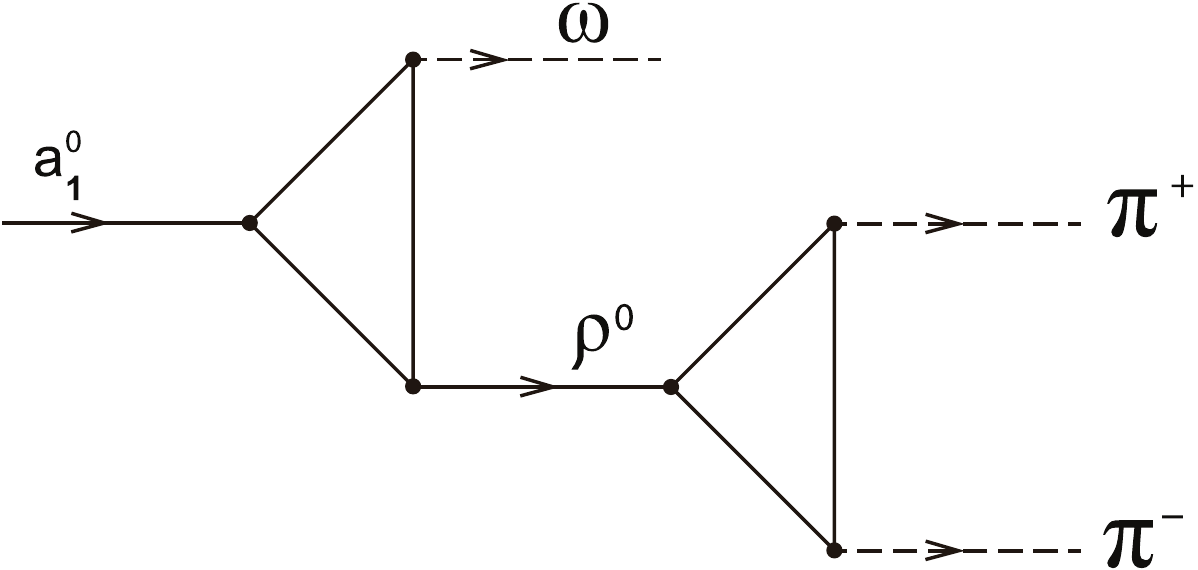}}
\caption{A typical Feynman diagram describing the $\rho^0$-exchange mode for the $a_{1}(1260) \to \omega \pi^{+} \pi^{-}$ decay.}
\label{fig4}      
\end{figure}

\subsection{The $\rho^0$ exchange mode}
\label{ss4.1}
On the theoretical side, the only difference between $f_{1}(1285) \to \rho^{0} \rho^{0}\to \rho^{0}\pi^{+} \pi^{-}$ and $a_{1}(1260) \to \omega\rho^{0}\to\omega \pi^{+} \pi^{-}$ decay amplitudes is the replacement of $f_1\rho^0\rho^0$ quark triangle by the $a_1^0\omega\rho^0$ one [compare Fig.~\ref{fig1} and Fig.~\ref{fig4}]. These vertices are originated by the same quark-loop diagram, including an overall factor which comes out from the isospin trace calculations. In the case of $a_1^0\omega\rho^0$ vertex we have tr[$(a_1^0\tau_3)(\omega\tau_0)(\rho^0\tau_3)$]=$2a_1^0\omega\rho^0$. That should be compared with tr[$(f_1\tau_0)(\rho^0\tau_3)(\rho^0\tau_3)$]=$2f_1\rho^0\rho^0$. Thus, for the channel (a) one can write immediately 
\begin{eqnarray}
F_1^{(\rho^0 )}&=& \left(\frac{\alpha_\rho^2}{2m^2 }\right)\frac{m_{a_1}^2+m_\omega^2 -2m_{a_1}(\varepsilon_\omega -\varepsilon_- )}{m_\rho^2 - s  },  \\
F_2^{(\rho^0 )}&=&-\left(\frac{\alpha_\rho^2}{2m^2 }\right)\frac{m_{a_1}^2+m_\omega^2-2m_{a_1}(\varepsilon_\omega-\varepsilon_+ )}{m_\rho^2 - s }, \\
F_3^{(\rho^0 )}&=&\left(\frac{2\alpha_\rho^2}{m^2 }\right)\frac{m_{a_1}^2+m_\omega^2-m_{a_1}\varepsilon_\omega}{m_\rho^2 - s  }, \\
F_4^{(\rho^0 )}&=&\left(\frac{-\alpha_\rho^2}{m^2 }\right)\frac{1}{m_\rho^2 - s  }, 
\end{eqnarray}
where $\varepsilon_\omega$ is the energy of the $\omega (782)$ meson in the rest frame of $a_1(1260)$ meson. In this reference frame, we have $s=m_\omega^2-m_{a_1}(2\epsilon_\omega -m_{a_1})$ [In the following, for simplicity, we put $m_\omega =m_\rho$.] This channel gives rather low value $\Gamma (a_{1} \to\omega\rho^0 \to \omega \pi^{+} \pi^{-})=12\,\mbox{keV}$.

\subsection{The $\rho^\pm$ exchange modes}
\label{ss4.2}
\begin{figure}
\resizebox{0.40\textwidth}{!}{%
 \includegraphics{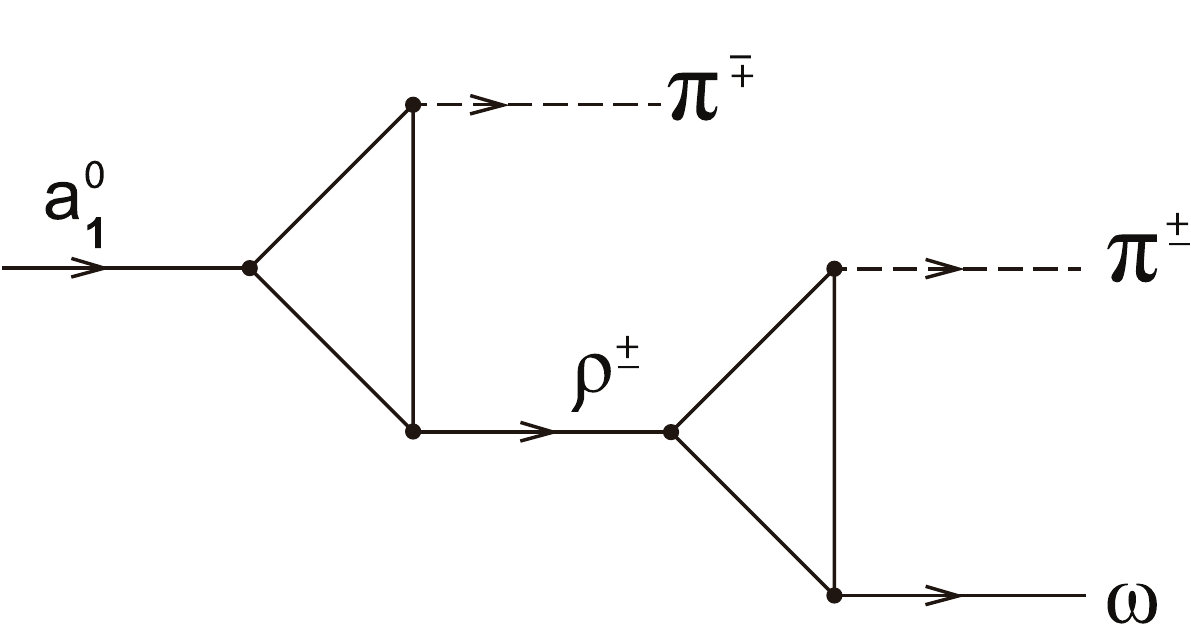}}
\caption{Two Feynman diagrams describing the $\rho^+$ and $\rho^-$ exchange modes for the $a_{1}(1260) \to \omega \pi^{+} \pi^{-}$ decay.}
\label{fig5}      
\end{figure}

The amplitude which describes the process shown in Fig.~\ref{fig5} is the analog of the $a_1^\pm$ exchange modes (b) for the $f_1\to\rho^0\pi^+\pi^-$ decay. Here, there is a common vertex $a_1\rho\pi$, where the $a_1(1260)$-meson is on-shell now 
\begin{equation}
{\cal L}_{a_1\pi\rho}^{a_1-mass}=i\left(\frac{2\kappa m^2}{f_\pi}\right)a_{1}^{0\mu}\left(\partial^\nu\rho^+_{\mu\nu}\pi^- -\partial^\nu\rho^-_{\mu\nu}\pi^+\right).
\end{equation}
Another vertex, $\rho\omega\pi$, which is responsible for the unnatural-parity decay process, is similar to the vertex $a_1f_1\pi$ [see Eq.~(\ref{fap})]. 
\begin{equation}
\label{rop}
{\cal L}_{\rho\omega\pi}=3g_{a} e^{\alpha\beta\mu\nu} \omega_{\nu} \partial_\beta\vec\rho_{\mu} \partial_\alpha\vec\pi ,
\end{equation} 
where a coupling constant $g_a$ is given by Eq.~(\ref{ga}).

From these Lagrangian densities we find the amplitude $T^{(\rho^\pm)}=T^{(\rho^+)}+T^{(\rho^-)}$ corresponding to the diagrams shown in Fig. \ref{fig5}   
\begin{eqnarray}
\label{rho+-exch}
T^{(\rho^\pm)}&=&ig_a \left(\frac{6\kappa m^2}{f_\pi}\right) e_{\mu\nu}^{\cdot\cdot\alpha\beta} \epsilon_\beta (l) \epsilon^*_\alpha (p) p^\mu p_+^\nu \nonumber \\
&\times&\frac{u}{m_{\rho}^2-u}-(p_+\leftrightarrow p_-).
\end{eqnarray}
This result differs from the one we had previously, considering the $a_1^\pm$ exchange contributions to the $f_1\to\rho^0\pi^+\pi^-$ amplitude. In particular, this amplitude vanishes if one makes a replacement $\epsilon^*_\alpha (p)\to p_\alpha$. Therefore the amplitude is a gauge invariant expression, and it is not possible to fix the ambiguity in $g_a$ by insisting that this symmetry is preserved [the transition to the radiative decay amplitude $a_1^0\to\gamma\pi^+\pi^-$ does not lead to any restrictions on the parameter $a$]. However, one can fix $a$ from the $f_1\to\rho^0\pi^+\pi^-$ decay, as we did in Sec.~\ref{ss3.4}. There we got $a=5/12$. In doing this, we also improve the description of $\rho^\pm\to\pi^\pm\gamma$ decay in the NJL model. Let us remind that the decay width of this process is given by
\begin{equation}
\label{rpg}
\Gamma (\rho^\pm\to\pi^\pm\gamma)=\frac{\alpha g_a^2}{96\pi\alpha_\rho}\left(\frac{m_\rho^2 -m_\pi^2}{m_\rho }\right)^3.
\end{equation}    
So, at $a=5/12$ we find that $\Gamma (\rho^\pm\to\pi^\pm\gamma)=78\,\mbox{keV}$. This is a little high compared to the experimental value $\Gamma (\rho^\pm\to\pi^\pm\gamma)=67.1\pm 7.4\,\mbox{keV}$ \cite{Agashe:2014kda} but is definitely better than $\Gamma (\rho^\pm\to\pi^\pm\gamma)=87\,\mbox{keV}$ obtained in \cite{Volkov:1986zb}. 
 
Finally, using Eq.~(\ref{rho+-exch}) and Eq.~(\ref{gafixed}), we come to the following form factors 
\begin{eqnarray}
&F_1^{(\rho^\pm)}&= \frac{3\alpha_\rho^2}{2m_\rho^2}\left(4-\kappa m^2\right) \frac{u}{m_{\rho}^2-u } \nonumber \\
&F_2^{(\rho^\pm)}&=-\frac{3\alpha_\rho^2}{2m_\rho^2}\left(4-\kappa m^2\right) \frac{t}{m_{\rho}^2-t }   \nonumber \\
&F_3^{(\rho^\pm )}&=F_1^{(\rho^\pm)}-F_2^{(\rho^\pm)}.
\end{eqnarray}
It gives $\Gamma (a_{1}(1260) \to \pi^\pm\rho^\mp\to\omega \pi^{+} \pi^{-})=517\,\mbox{keV}$.

\begin{figure}
\resizebox{0.30\textwidth}{!}{%
 \includegraphics{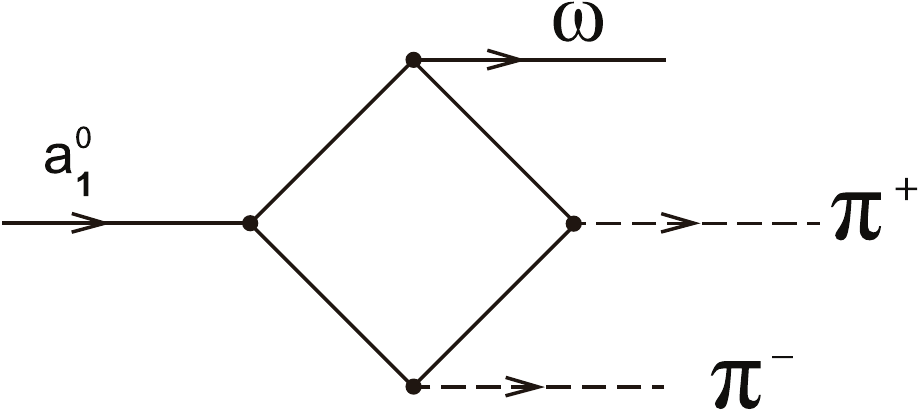}}
\caption{The box Feynman diagram describing the direct mode for the $a_{1}(1260) \to \omega \pi^{+} \pi^{-}$ decay. We do not show the diagrams which can be obtained by permuting the final states.}
\label{fig6}      
\end{figure}

\subsection{The box diagrams}
In Fig.~\ref{fig6} there is drawn a typical diagram that describes the direct decay mode. It depicts the process where pions interact with quarks without derivative $\bar q \gamma_5\vec\tau\vec\pi q$. There are also diagrams which include the derivative coupling of pions with quarks $\bar q\gamma^\mu \gamma_5 \partial_\mu \vec\pi \vec\tau q$. In the corresponding amplitude (\ref{boxa1}), the contribution of each coupling with a derivative is proportional to $\kappa m^2$. We also do not show the diagrams which can be obtained by permuting the final states, although we take them into account. The result of calculations of all box diagrams in the leading order of derivative expansion is    
\begin{eqnarray}
\label{boxa1}
T^{(d)}&=&i\frac{\alpha_\rho N_c}{2\pi f_\pi^2} e^{\mu\nu\alpha\beta}\epsilon_\beta (l)\epsilon_\alpha^*(p)\left[(1-2\kappa m^2)p_\mu q_\nu + \right. \nonumber \\
&+& \left. (\kappa m^2)^2 p_+^\mu p_-^\nu \right].
\end{eqnarray} 

The corresponding form factors are
\begin{eqnarray}
F_1^{(d)}&=&-F_2^{(d)}=\frac{\alpha_\rho N_c}{2\pi f_\pi^2}\left(1-2\kappa m^2\right), \nonumber \\
F_3^{(d)}&=&\frac{\alpha_\rho N_c}{\pi f_\pi^2}\left(1-2\kappa m^2+\frac{1}{2}(\kappa m^2)^2\right).
\end{eqnarray}
It follows then that $\Gamma (a_{1}(1260) \to \omega \pi^{+} \pi^{-})_{box}=52\,\mbox{keV}$.

As we already know from Sec.~\ref{ss3.4}, the last term in (\ref{boxa1}) can be a source of the gauge symmetry breaking [through the VMD mechanism]. We have checked gauge invariance for the $a_1\to \gamma\pi^+\pi^-$ decay amplitude. This symmetry is protected by contributions, which are not generated by the VMD mechanism. The details will be given in the separate paper.

\subsection{The $a_{1}(1260) \to \omega \pi^{+} \pi^{-}$ decay width}
\label{ss4.4}
 We have already shown that diagrams in Fig.~\ref{fig5} yield the dominant contribution to the $a_{1}(1260) \to \omega \pi^{+} \pi^{-}$ decay width. Our aim now is to clarify the interference effects.  

Let us consider first the sum of diagrams plotted in Figs. \ref{fig5} and \ref{fig6}. The corresponding form factors can be combined in the following structures
\begin{eqnarray}
&&F_1^{(\rho^\pm)}+F_1^{(d)}=\left(\frac{3\alpha_\rho^2}{2}\right)\frac{4-\kappa m^2}{m_\rho^2-u}+3F_1^{(c)},\nonumber \\
&&F_2^{(\rho^\pm)}+F_2^{(d)}= -\left(\frac{3\alpha_\rho^2}{2}\right)\frac{4-\kappa m^2}{m_\rho^2-t}-3F_1^{(c)},\nonumber \\
&&F_3^{(\rho^\pm)}+F_3^{(d)}=\frac{3\alpha_\rho}{\pi f_\pi^2}\left[1-4\kappa m^2+(\kappa m^2)^2\right] \nonumber \\
&&+ \left(\frac{3\alpha_\rho^2}{2}\right)(4-\kappa m^2)\left(\frac{1}{m_\rho^2-u}+\frac{1}{m_\rho^2-t}\right),
\end{eqnarray}
where $F_1^{(c)}$ is given by (\ref{contf}). From that we deduce that there is destructive interference between the amplitudes arising from these two channels. As a result, their contribution to the decay width turns out to be essentially suppressed $\Gamma^{(\rho^\pm)+(d)}=(517+52-326)\,\mbox{keV}=243\,\mbox{keV}$. 

Additionally, a large suppression occurs due to destructive interference between the $\rho^0$-exchange amplitude of Fig.~\ref{fig4} and the sum of diagrams shown in Figs.~\ref{fig5} and \ref{fig6}. This leads to a rather low value $\Gamma(a_{1}\to \omega \pi^{+} \pi^{-}) =(12+243-168)\,\textrm{keV}= 87 \textrm{ keV}$.

\section{Conclusions}
\label{concl}
The purpose of this paper has been to use our knowledge of the structure of the triangle quark $f_1\rho\gamma$ anomaly for studying $f_1(1285)\to\rho\pi\pi$ and $a_1(1260)\to\omega \pi\pi$ anomalous decays, where similar vertices $f_1\rho\rho$ and $a_1\rho\omega$ arise as a part of more sophisticated chiral dynamics. As a result, it has been found that theoretical estimation for the $f_1\to\rho\pi\pi$ decay width [$\Gamma(f_{1}\to \rho^{0} \pi^{+} \pi^{-}) = 2.78 \textrm{ MeV}$] agrees well with the experimental value (\ref{expf1}). It has been also obtained [for the first time] a theoretical prediction for the rate of the $a_1\to\omega \pi\pi$ decay, $\Gamma(a_{1}\to \omega \pi^{+} \pi^{-}) = 87 \textrm{ keV}$. 

Both processes receive contributions from the box AAAV anomaly which is carefully calculated here. This anomaly is less studied experimentally and can be an interesting subject for future investigations. Our calculations indicate clearly that there is a large interference between box and triangle anomalies. The strong suppression of the $a_1\to\omega \pi\pi$ decay found in our work is a direct consequence of such destructive interference. It would be informative to measure the rate of this decay. From this one could learn about the structure of the $a_1(1260)$ state. If the experimental result will support the value found in this paper, one can conclude that $q\bar q$ component is dominated in $a_1(1260)$. If not, it will reinforce the idea of the dynamical, or molecular, nature of the $a_1(1260)$ meson \cite{Lutz04,Oset05,Hosaka11}. In fairness it has to be said that the internal structure of the $f_1(1285)$ meson is also not well understood. Thus, the obtained agreement with the experimental result for its decay ratio is a significant and non-trivial argument in favour of $q\bar q$ content of $f_1(1285)$.

Our estimates are based on the local vertices of the effective meson Lagrangian of the NJL model where meson states are treated as the nearly stable quark-antiquark particles. Following the idea of $1/N_c$ expansion we assumed that in the long-wavelength regime only local contributions with minimal number of derivatives are important. We suppose that the qualitative and quantitative features that emerge in our simplified consideration would persist in a more accurate calculation. This can be done in the future as soon as new empirical data will be available. Nonetheless, the undoubted merit of the described results, as compared to the already known ones in the literature, is that they are relied on the more detailed dynamical picture.

\end{document}